\documentclass[sigconf]{nimeart}
\AtBeginDocument{%
  }

\setcopyright{cc}
\copyrightyear{2026}
\acmYear{2026}
\acmDOI{}

\usepackage{textcomp}

\acmConference[NIME '26]{International Conference on New Interfaces for Musical Expression}{June 23--26,
  2026}{London, UK}

\acmISBN{}

\usepackage{enumitem}

\begin{document}

\title{Exploring Breathing-Music Coupling: Using the \textit{Breathing Mirror} for Somatic Reflection in Piano Performance}

\author{Ziyue Piao}
\orcid{0009-0007-1281-9469}
\affiliation{%
  \institution{IDMIL, MPBL, CIRMMT, McGill University}
  \city{Montreal}
  \state{QC}
  \country{Canada}
}
\email{ziyue.piao@mail.mcgill.ca}

\author{Yohei Wada}
\affiliation{%
  \institution{Yamaha Corporation}
  \city{Hamamatsu}
  \country{Japan}}
\email{yohei.wada@music.yamaha.com}

\author{Isabelle Cossette}
\affiliation{%
  \institution{MPBL, CIRMMT, McGill University}
  \city{Montreal}
  \state{QC}
  \country{Canada}
}
\email{isabelle.cossette1@mcgill.ca}

\author{Marcelo M. Wanderley}
\affiliation{%
  \institution{IDMIL, CIRMMT, McGill University}
  \city{Montreal}
  \state{QC}
  \country{Canada}
}
\email{marcelo.wanderley@mcgill.ca}

\author{Akira Maezawa}
\affiliation{%
  \institution{Yamaha Corporation}
  \city{Hamamatsu}
  \country{Japan}}
\email{akira.maezawa@music.yamaha.com}

\renewcommand{\shortauthors}{Piao et al.}

\begin{abstract}

    While breathing is essential to living and for sound production in some instruments, for pianists, it is often a hidden and automatic process, making it difficult to analyze or refine. A critical gap exists between data and awareness: while sensors record precise physical metrics, they fail to capture the performer's somatic experience. Conversely, the high cognitive load of performance makes it nearly impossible for musicians to recall their internal states with temporal precision. To address this, we present a system, \textit{Breathing Mirror}, and associated methodology designed to externalize the pianist's internal somatic experience through three analytical lenses: a Baseline View (synchronized signals), a First-Person View (subjective recall), and an Interpersonal View (collaborative reflection).

    Through a four-week longitudinal study with a skilled amateur pianist (35 years of experience), we evaluated the system's effectiveness by recording respiratory data using textile-integrated strain sensor belts. The results show that the \textit{Breathing Mirror} reveals some patterns of breathing-music coupling and identifies critical blind spots where objective data diverges from subjective perception. Furthermore, we propose four somatic themes regarding the link between breathing and musical elements, offering a foundation for future large-scale validation across a broader range of pianists. This work provides a new way to study body signals, transforming breathing from an internal biological function into an articulate expressive parameter.

\end{abstract}



\maketitle

\section{Introduction}

Breathing is essential for life and serves as an expressive tool that influences musical phrasing, timing, and emotional communication. While breathing is necessary for making sound in wind instruments and voice, pianists do not rely on breath to create tone~\cite{king2017supporting}. Still, the respiratory patterns of pianists present that even without a direct physical connection to sound production, pianists need to be aware of their breathing in both teaching and performance settings. Recent studies have shown links between pianists' breathing cycles and musical structures, which suggests that breathing connects physical gestures to musical structure~\cite{king2017supporting, sakaguchi2016relationship, laczika2013flauto}.

Building on Merleau-Ponty's notion of the body as an `inconceivable absence,' we recognize that while our physical presence is constant, it remains largely transparent to our conscious mind~\cite{merleau2014phenomenology}. Sensors can record physical metrics, but they often fail to capture the subjective feelings and artistic intentions behind the movement. Conversely, first-person methods like somaesthetic reflection can document somatic experiences~\cite{hook2018designing}, yet the high cognitive demand of complex piano performance makes it nearly impossible to recall internal physiological states within the rapid musical timeline. This inability to consciously monitor the body during complex tasks leads to a kinesthetic mismatch, a discrepancy where the brain's internal representation of the performance diverges from the body's actual physiological state~\cite{Piao2026Reconciling}. Such a mismatch obscures the performer's ability to reflect on somatic nuances with precision and limits the researcher's understanding of how these bodily states shape performance. 

In this paper, we present the \textit{Breathing Mirror}, a visualization interface designed to bridge the somatic gap by integrating multimodal respiratory data into a somatic reflective tool. Our approach to somatic reflection is informed by Rigg's concept of somatic learning, which treats the body as a site of lived experience rather than a biological machine. We focus on the transition where pre-reflective bodily awareness is brought into conscious thought and articulated as somaesthetic knowledge~\cite{rigg2018somatic}. By enabling performers and researchers to co-observe respiratory signals alongside musical output, the interface facilitates a ``deep listening'' to the body, allowing the performer to transform pre-reflective bodily sensations into articulated musical insights. To evaluate this approach, we conducted a month-long longitudinal study with a skilled amateur pianist (35 years of experience), reflecting their synchronized breathing signals and musical output during piano performance. 

The main contributions of this work are: 

\begin{itemize} [topsep=3pt, itemsep=2pt, parsep=0pt, partopsep=0pt, leftmargin=15pt]
    \item We introduce the \textit{Breathing Mirror} system and a methodology that integrates a Baseline View, a First-Person View, and an Interpersonal View for structured somatic reflection. 
    \item We propose four somatic themes regarding the generalizability of the breathing-music coupling for further evaluation with a broader cohort of expert pianists. 
\end{itemize}

\section{Related Work}

\subsection{Phenomenological Foundations}
Bodily reflection draws on Edmund Husserl's theories of reflection and Merleau-Ponty's phenomenological philosophy~\cite{merleau2014phenomenology, moran2010husserl}, positioning the body as a site of conscious inquiry rather than a mere instrument. In dance performance, Buttingsrud defines embodied reflection as a state of attentive presence where the performer actively engages with their bodily sensations~\cite{buttingsrud2021bodies}. Bodily reflection techniques now represent a growing interdisciplinary area that bridges phenomenology, embodied cognition, and performance practice across diverse performance domains~\cite{foultier2023letting}. While traditional models of expertise emphasize automaticity and the body's inconspicuousness, recent research suggests that expert performers strategically deploy conscious attention to facilitate continuous improvement and artistic expression~\cite{buttingsrud2021bodies, foultier2023letting}.

In music, Reybrouck et al. advocate for a ``centripetal'' approach, emphasizing the cyclical relationship between bodily action and sound production~\cite{reybrouck2024music}. Within this framework, somatic reflection is seen as integral to musical learning, and this aligns with Leman's concept of embodiment, where the body acts as a mediator between musical intention and physical execution~\cite{leman2015role, leman2007embodied}. 
Musicians have also widely adopted practical approaches such as the Alexander Technique to enhance somatic awareness and physical coordination~\cite{buckoke2013alexander, nicholls2014body, cotik2019concepts}. For example, video recording a performance and analyzing the movement can be a helpful learning tool~\cite{cotik2019concepts}. However, a significant challenge arises in complex musical performances: as the performer's focus shifts toward musical output, they often experience a kinesthetic mismatch. This phenomenon describes the discrepancy between the brain's internal representation of an action, the `felt' intent, and the actual physiological state of the body~\cite{proske2012proprioceptive, schmidt2018motor, Piao2026Reconciling}. Under high cognitive demand, the body sometimes becomes nearly impossible for musicians to maintain a sustained awareness of foundational states like breathing, despite their critical role in performance.



\subsection{Somatic Interaction Tools for Performance}

Soma design, pioneered by Kristina Höök~\cite{hook2018designing}, centers the lived, feeling body as both the subject and medium of design. Within the NIME community, this paradigm shifts focus from technically-driven approaches to the performer's bodily experience as the primary site of meaning-making~\cite{martinez2020soma}. 
Projects such as Body Electric~\cite{cotton2021body}, DogDog~\cite{bigoni2020dogdog}, and Suspended Circles~\cite{bang2023suspended} demonstrate how soma-based inquiry can generate novel sensing mechanisms and aesthetic experiences deeply rooted in the performer's feeling.

Because somatic experiences are often subtle and pre-reflective, researchers have applied various methods to externalize and articulate these internal sensations.
Autoethnography allows designers of novel gestural systems to use somatic methods as a foundation for critically reviewing their own performance experiences~\cite{mainsbridge2022feeling}. To bridge the gap between action and awareness, researchers have utilized micro-phenomenology to help performers describe tiny and hidden sensations, such as subtle tactile contact with an instrument, typically passed unnoticed during musical performance~\cite{reed2022exploring}.

However, many of these approaches are adapted from somatic practices like Feldenkrais~\cite{hillier2015effectiveness, martinez2020soma}, which prioritize slow, deliberate movement to refine proprioceptive acuity. While effective for training, these methods often lack the temporal resolution necessary for the rapid dynamics of music. Furthermore, they are difficult to implement during live performance, where the intense cognitive load prevents musicians from consciously accessing or articulating their internal states in real-time.

\subsection{Breathing and Body in Piano Performance}
Breathing is a particularly rich domain for soma design, as it is an automatic bodily function and a movement we can consciously control~\cite{cotton2021body, tapparo2022bodily}. We define \textit{Breathing-Music Coupling} as the correlation between respiratory signals, spanning somatic, kinesthetic, and sensor data, and the structural or expressive parameters of the musical information. While recent NIME research has leveraged breathing-aware interfaces, these applications have primarily focused on vocal or wind contexts where respiration is a mechanical necessity for sound production~\cite{cotton2021body, freire2024body, kilic2023corsetto, piao2022sensing, li2021wearable}. In such domains, somatic sensing tools allow performers to externalize internal states, facilitating deeper somaesthetic appreciation and shared felt experiences.

In piano pedagogy and performance, however, this somatic dimension remains largely underexplored. Traditional piano learning interfaces primarily emphasize external correctness, such as pitch accuracy and rhythmic timing~\cite{liu2023pianosyncar, lee2026fixical}. While interactive systems like MirrorFugue III~\cite{xiao2013mirrorfugue} use visual projections to examine the relationship between the physical body and recorded performances, their focus remains on external gestural reflection. Existing interfaces often neglect the internal physiological dimensions of piano performance, failing to account for how respiratory patterns actively shape and support a pianist's expressive intent.

\section{The \textit{Breathing Mirror} Interface}
To capture the nuanced interplay between breath and performance, we designed the \textit{Breathing Mirror} interface that supports interactive data visualization and time-series annotation for post-hoc analysis. Rather than employing real-time visualization, we utilized pre-recorded data to avoid the effort of monitoring one's breath, along with competing with the mental effort required for complex finger movements or musical grouping. The interface displays respiratory signals from two high-fidelity wearable sensor belts to provide a visual-analytical environment, enabling the precise temporal alignment of respiratory signals with musical performance data.

\subsection{The Breathing Belts}

\begin{figure}[H]
  \centering
  \includegraphics[width=\linewidth]{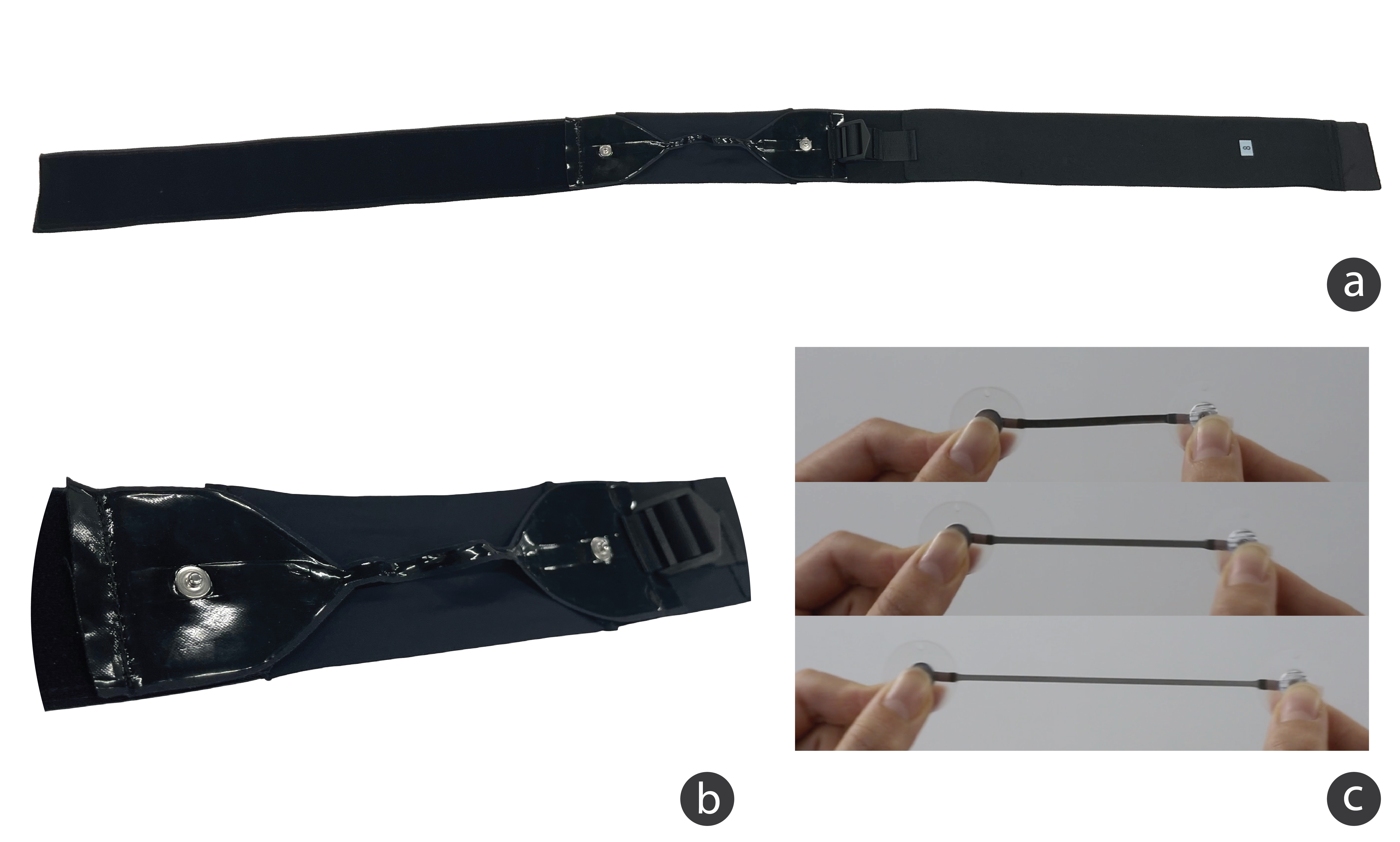}
  \caption{The breathing belt for respiratory measurement. (a) Overview of the belt assembly, featuring an embedded strain sensor at the center; (b) detailed close-up of the sensor placement; (c) visualization of sensor stretching across low, medium, and high resistance states.}
  \Description{The figure illustrates the breathing belt for respiratory measurement. (a) Overview of the belt assembly, featuring an embedded strain sensor at the center; (b) detailed close-up of the sensor placement; (c) visualization of sensor stretching across low, medium, and high resistance states.}
  \label{fig:gummibelt}
\end{figure}

The sensor utilized in this study is a carbon-nanotube (CNT)-based strain sensor (GummiStra\texttrademark, Yamaha Corporation, Hamamatsu, Japan) developed by Suzuki et al.~\cite{suzuki2016rapid} characterized by its flexibility, lightweight profile, and comfort. The material is ideally suited for wearable applications, particularly for textile-based, real-time human motion sensing. Breathing belts embedded with the sensors have already demonstrated effectiveness in both medical contexts and wind instrument monitoring~\cite{kobayashi2021use, shoji2025measurement}. The sensor and its accompanying breathing belt are demonstrated in Figure~\ref{fig:gummibelt}.

Compared to existing respiratory monitoring technologies in NIME research, such as traditional strain gauges~\cite{piao2022sensing} or the conductive rubber-based ATEM-P sensor~\cite{jones2023wearable}, the sensor offers several advantages:

\begin{enumerate}
    \item \textbf{Linearity:} The material provides a more linear relationship between stretch and resistance, facilitating data interpretation without the need for complex post-processing or intensive normalization.
    \item \textbf{Integration:} As a textile-integrated sensor utilizing thin-film CNT coatings, it is significantly thinner and more pliable than bulkier alternatives. This allows for seamless embedding within garments, expanding its applicability across diverse musical performance contexts.
    \item \textbf{Transparency:} It can be engineered with a lower spring constant. Consequently, the sensor deforms with the body without providing significant counter-resistance, allowing the musician's gestures to remain undisturbed.
\end{enumerate}

\subsection{Data Acquisition and Transmission}
To ensure high-fidelity synchronization between the piano performance and the performer's respiratory responses, we utilized a multimodal recording system. Musical data was captured via a Yamaha Stage Piano CP88, which recorded MIDI events with a temporal resolution of roughly 1 ms. Simultaneously, the acoustic performance was captured through the instrument's internal audio interface at a sampling rate of 44.1 kHz, synchronized by a common clock signal.

Respiratory signals were captured using two breathing belts. Following the classical two-compartment model for measuring breathing~\cite{konno1967measurement}, the two belts were positioned at the mid-thorax (lower rib cage) and the umbilical level (abdomen) to independently monitor thoracic and abdominal displacement. These sensors were interfaced with a National Instruments USB-DAQ microcontroller, which sampled the analog signals at 10 kHz. Breathing signals transmission to the local workstation was facilitated via a high-speed Wired Serial USB with NI daqmx protocol connection, ensuring the stable throughput and low latency required for real-time visualization.

The MIDI, audio, and respiratory data streams were synchronized using Rosendahl mif4. This protocol allowed for the generation of unified, time-stamped files. MIDI data was preserved in .mid format, while the audio and respiratory signals were encoded as .wav files to maintain high-resolution waveforms for subsequent thematic analysis and somatic reflection. Finally, the multimodal data were transferred via a local shared network directory to a 13-inch MacBook Pro for visualization within the \textit{Breathing Mirror} interface.

\subsection{The Visualization Interface}
\label{subsec:interface}

\begin{figure*}[h]
  \centering
  \includegraphics[width=0.8\linewidth]{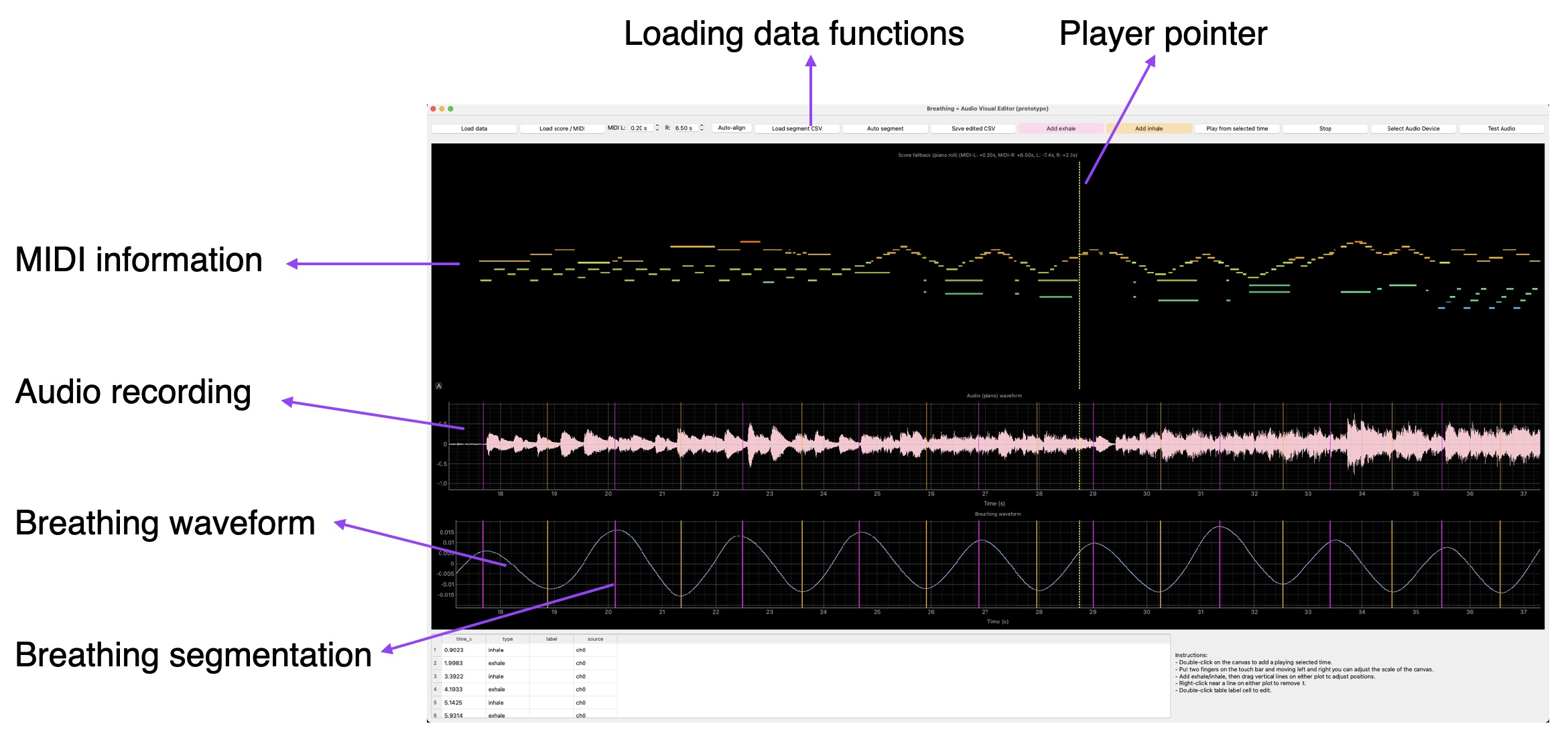}
  \caption{The multi-modal visualization interface. The display integrates real-time MIDI data, audio waveforms, and respiratory signals to provide a comprehensive overview of the performance.}
  \Description{The multi-modal visualization interface. The display integrates real-time MIDI data, audio waveforms, and respiratory signals to provide a comprehensive overview of the performance.}
  \label{fig:interface}
\end{figure*}

Figure~\ref{fig:interface} shows the interface acting as a visual mirror that recalls physiological dynamics alongside musical performance. The main functions are as follows:

\begin{enumerate}
    \item \textbf{Multimodal Canvas:} The interface displays synchronized MIDI, audio, and respiratory data upon loading. 
    \item \textbf{Dynamic Annotation:} The system automatically segments breathing into inspiration (yellow) and expiration (pink) phases. Users can interactively correct these markers by adding, moving, or deleting bars. 
    \item \textbf{Reflective Playback:} A double-click on the canvas initiates time-aligned playback of both audio and visual data. This allows performers to directly compare their felt sensations during the performance with the recorded physiological evidence.
    \item \textbf{Synchronized Qualitative Table:} To facilitate post-hoc analysis, the interface includes an annotation table linked to the timeline. Clicking a breathing segment automatically locates its entry in the table, where pre-populated topic labels serve as a reference for further analysis by mapping subjective somatic reflections to precise timestamps.
\end{enumerate}

We adopted a method similar to the standardized ratio established by Banzett et al.~\cite{banzett1995simple}, in which thoracic signals are weighted twice as heavily as abdominal signals to estimate respiratory volume. While this weighting traditionally relates to surface area and volume changes, we use the thoracic channel as the primary reference for both visualization and the temporal onset detection described in section~\ref{subsec:algorithm}.

\subsection{Breathing Segmentation Algorithm}
\label{subsec:algorithm}

The respiratory segmentation pipeline extracts breathing phases from wearable sensor data through three primary stages: filtering, peak detection, and multi-channel fusion. Specifically, we implement a multi-channel fusion stage to address instances where thoracic and abdominal signals exhibit temporal misalignment or phase shifts. To mitigate motion artifacts and subtle baseline drift resulting from an inherent byproduct of gradual sensor displacement during hours of recording sessions, the raw signal $x[n]$ is processed using a fifth-order Butterworth band-pass filter ($0.05$--$2.0$ Hz). We employ a zero-phase filtering approach via forward-backward filtering to ensure that respiratory onsets remain aligned with the musical events. The resulting signal is smoothed using a moving average window ($W=15$) to produce $y_{smooth}[n]$.

The inhalation and exhalation events are then identified using an adaptive prominence threshold $\Gamma$. This threshold is dynamically calculated based on the signal's standard deviation $\sigma(y_{smooth})$, allowing the system to remain robust across varying breathing depths:
\begin{equation}
\Gamma = \max(0.05, 0.2 \times \sigma(y_{smooth}))
\end{equation}

To ensure high-fidelity event detection, a fusion logic integrates the thoracic ($ch_0$) and abdominal ($ch_1$) sensor streams. We prioritize the thoracic channel for onset detection, a decision supported by our initial pilot testing and empirical observations, which show that the thoracic signal exhibited a higher Signal-to-Noise Ratio compared to the abdominal channel. We define $t_{ch0}$ and $t_{ch1}$ as the detected timestamps of a respiratory peak (or onset) in the respective channels. Candidate event $E(t)$ detect in $ch_0$ is validated if it satisfies at least one of the following criteria:
\begin{itemize}
    \item \textbf{Temporal Coherence:} A candidate event $t_{ch0}$ is cross-validated if a corresponding abdominal peak $t_{ch1}$ is identified within a defined temporal window. Specifically, the event is validated if $\exists \, t_{ch1} : |t_{ch0} - t_{ch1}| \le \tau$, where the tolerance $\tau$ is set to $0.75$~s to account for physiological phase shifts between thoracic and abdominal expansion.
    
    \item \textbf{Signal Confidence:} To preserve high-intensity respiratory gestures that may be isolated to the upper torso, an event is also validated if its relative amplitude $|A_{ch0}|$ exceeds a normalized confidence threshold $\eta \ge 0.7$. Here, $\eta$ represents a scaling factor of the global standard deviation $\sigma(y_{smooth})$, ensuring the algorithm adapts to the specific dynamic range of the performance.
\end{itemize}

Formally:
\begin{equation}
E(t_{ch0}) =
    \begin{cases}
        \text{Valid} & \text{if } \exists  t_{ch1} : |t_{ch0} - t_{ch1}| \le \tau \\
        \text{Valid} & \text{if } \frac{|A_{ch0}|}{\sigma(y_{smooth})} \ge \eta \\
        \text{Invalid} & \text{otherwise}
    \end{cases}
\end{equation}

To filter out noise from the pianist's body movements, we exclude any brief ``trough-peak-trough'' signal patterns shorter than 0.6 s.

While this pipeline provides an automated initial pass, the interface prioritizes data fidelity for qualitative analysis. Consequently, the algorithm serves as an assisted-segmentation tool. Although all detected events are manually refined by the researchers and verified by the pianist, this approach significantly reduces the time required compared to manual labeling from scratch. Future iterations will continue optimizing the automated model.

\section{Investigation with the Practitioner}

\subsection{Methodology}
We focus on using somatic reflection to externalize the performer's pre-reflective respiratory gestures. To ensure analytical rigor, we adopt an idiographic approach, prioritizing a deep investigation of a single somatic expert's lived experience over broad statistical sampling~\cite{smith2021interpretative}. While this section outlines the conceptual lenses, the specific experimental procedures and data acquisition protocols are detailed in Section~\ref{subsec:procedure}. In summary, the \textit{Breathing Mirror} does not aim to provide an objective truth for the performer to rationalize; rather, it serves as a probe that triggers kinesthetic recall. By re-experiencing the temporal alignment of breath and music, the performer transitions from mere observation to somatic re-living of their performance.

\subsubsection{The Baseline View: Data-supported Objective Reference}
The baseline view functions as a methodological anchor, providing a physiological record to support reflection. Its role is to transform ephemeral bodily sensations into a synchronized, time-stamped representation of the performance. By externalizing respiratory signals alongside musical data, this view acts as a navigational map to support a first-person and interpersonal view of reflection. It allows the performer or researchers to revisit specific moments of the performance with objective evidence, helping to resolve discrepancies between their subjective memory and their actual physical actions.

\subsubsection{The First-Person View: Preserving Somatic Authority}
The first-person view prioritizes the lived body, ensuring the musician's internal experience remains the primary source of truth. Unlike the other perspectives, this view relies exclusively on the performer's ``felt sense'' during the act of music-making. During the initial performance phase, the pianist plays in a familiar ``comfort zone'' to encourage a natural state of embodied expression, capturing the subjective qualities of the breath before any external data is displayed. In the subsequent reflection phase, this view is maintained by asking the performer to recall and articulate their sensations based on internal memory. 

\subsubsection{The Interpersonal View: Intersubjective Reflection}
The interpersonal view introduces a social and collaborative dimension to the somatic reflection process. In this view, the researchers and the performer jointly review the observable evidence from the music and breathing data. The researchers act as facilitators, identifying special signal patterns or structural shifts on the multimodal data to prompt deeper reflection through targeted questions. For example, when a sudden peak in the respiratory data aligns with a specific musical passage, the researcher employs targeted questioning like ``Looking at this peak in the respiratory data, was this an intentional breath you were aware of, or did it feel like a spontaneous bodily response to the music? How did you perceive the connection between your breath and the musical phrase in that moment?'' This joint attention fosters an intersubjective dialogue, allowing for a richer observation of the embodied musical experience.

\subsection{The Collaborative Practitioner Pianist}
While many piano traditions acknowledge breathing, the ability to articulate its role in performance consciously requires an exceptional level of somatic awareness. For this study, we collaborated with YK, a skilled amateur pianist with 35 years of classical training.~\footnote{The participant is classified as a skilled amateur pianist because her training was conducted through private instruction for personal enrichment rather than a formal performance degree. Furthermore, she has no history of professional or commercial musical engagements.} YK was selected as a somatic expert based on her unique pedagogical experience: her background spans 20 years of classical training with little emphasis on breathing, followed by 15 years of intensive, breath-centered practice. This rare dual-perspective makes YK uniquely situated to identify the nuances of respiratory patterns that are often pre-reflective for other performers. Her specialized lived experience serves as the primary driver for both the iterative refinement of the interface and the formalization of our somatic reflection procedures. The goal of this intervention was not pedagogical; the participant had already mastered the piece for concert performance. Rather, the study used the repertoire as a stable baseline to investigate the somatic gap between the performer's perceived breathing and her actual execution, using the interface as a mirror for deep somatic reflection.

\subsection{Procedure}
\label{subsec:procedure}

We conducted four interview sessions in a month, once per week. Session one served primarily as a familiarization phase to establish mutual understanding between the pianist and the researchers. Sessions two through four encounter with the system informed both the technical refinement of the interface and our theoretical understanding of the breathing-music coupling. At the end of sessions two through four, we exported the annotation notes from the interface and saved the audio recordings of the interviews for subsequent thematic analysis.

The first and second authors facilitated the sessions to ensure a multi-perspective observation. The second author acted as the primary interviewer, managing the \textit{Breathing Mirror} interface, replaying audio segments, and documenting somatic annotations in real-time. Complementing this role, the first author acted as a co-facilitator and observer, monitoring the respiratory patterns and adding questions to prompt deeper somatic reflection when significant breathing events were missing by either the second author or the pianist.

\subsubsection{Session 1: Introduction and System Familiarization (30 minutes)}
During the initial online meetup, two researchers and the pianist established a baseline for the study. The goals were to document YK's musical experience, identify a suitable piece for breathing analysis, and introduce the \textit{Breathing Mirror} system to prepare the participant for the upcoming sessions. 

The musical piece, the opening section of Nausicaä of the Valley of the Wind by Joe Hisaishi~\cite{hisaishi1984nausicaa}, is a three-minute contemporary work that was mutually selected for its explicit structural segmentation, featuring six sections that alternate between dense rhythmic ostinatos and expansive lyrical melodies. This stylistic contrast was designed to elicit a broad range of respiratory responses, allowing us to compare patterns during high-intensity technical passages against moments of rubato (expressive rhythmic freedom). As the pianist had previously received explicit instruction from her piano teacher on coordinating respiration with musical phrasing, she had already internalized the work's respiratory demands. This enabled the study to focus on her pre-reflective somatic habits and the embodied phrasing in fluent performance.


\subsubsection{Session 2: Performance Recording and Overview Annotation (120 minutes)}
\begin{figure}[H]
  \centering
  \includegraphics[width=\linewidth]{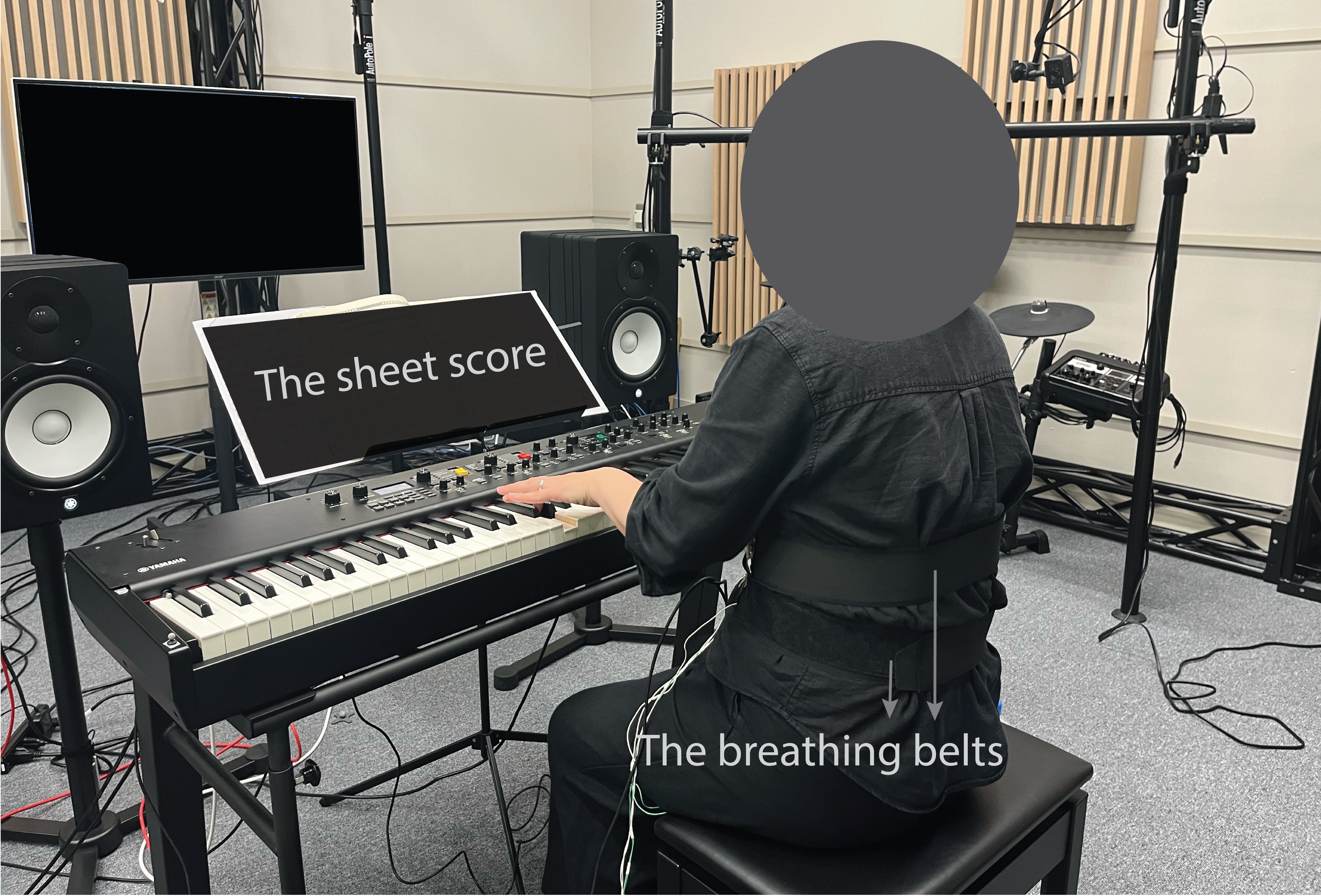}
  \caption{Experimental setup for session 2's multimodal data acquisition.}
  \Description{Experimental setup for session 2's multimodal data acquisition.}
  \label{fig:session2setup}
\end{figure}

The first 60 minutes of the second session were dedicated to the acquisition of high-quality multimodal data, capturing synchronized respiratory and musical signals using the experimental setup illustrated in Figure~\ref{fig:session2setup}. The procedure involved an initial warm-up period and multiple practice runs of the piece, followed by 2-3 recorded takes. This process continued until the performer was satisfied with a final take that they felt best represented their intended musical expression, particularly as shaped by their breathing.

The subsequent 60 minutes focused on overall annotation using \textit{Breathing Mirror}, which was also the participant's first experience with the interface. To facilitate deeper somatic recall, the pianist YK remained seated at the piano, physically touching the relevant musical phrases on the keys while watching the visualization interface. This multi-modal approach allowed YK to bridge her tactile memory with the digital data, resulting in nuanced breathing-music coupling annotations. For instance, the pianist identified a `big inhale' specifically triggered by a chord change from B to C at the 38.7-second mark, which was then logged in the interface as a combined structural and physiological marker: ``change of chord, big inhale B-C.'' 

YK was asked to review the performance data by systematically examining each recorded respiratory event, reflecting on her somatic experience, and articulating the relationship between these patterns and the musical structure. During this session, the practitioner took the lead, focusing primarily on the somatic reflection and annotation of the piece's first half, while the interviewer provided facilitative prompts to explore significant or anomalous observations. Finally, we tried to gather the pianist's feedback on the interface, using these insights to perform iterative usability refinements before the next session. 

\subsubsection{Session 3: In-Depth Annotation (60 minutes)}
This session took place one week later, utilizing a version of the interface optimized for high-resolution temporal navigation. Based on the practitioner's initial feedback, this update enabled precise timestamp labeling, allowing the performer to map specific somatic sensations to exact musical moments. In this session, the pianist was asked to re-examine the annotations of the first half of the piece from Session 2 to engage in a micro-analysis of the respiratory data. The objective was not merely to repeat the previous task, but to use the refined interface to identify nuanced breathing patterns, such as subtle body engagements or micro-pauses, that may have been overlooked during the broad overview of the first pass. This iterative review allowed the performer to differentiate between spontaneous somatic responses and those specifically influenced by prior pedagogical instruction.

Unlike the previous session, YK chose to engage directly with the interface without sitting at the piano, noting that this analytical distance better facilitated the detailed annotation process while still allowing for deep somatic reflection. By the conclusion of the session, we had completed a comprehensive micro-review of the first half of the selected piece. We initiated the same detailed reflection process for the second half.

\subsubsection{Session 4: In-Depth Annotation and Reflective Discussion (60 minutes)}
Following the protocol of the previous session, Session 4 began with the pianist re-evaluating the annotations from Session 3. The objective was to identify distinctive respiratory patterns, specifically those shaped by instructed patterns (e.g., an intentional expansion of the inhalation in preparation for a new musical section, guided by the teacher's instruction) or those that had previously remained unacknowledged (e.g., an unexpected, sharp respiratory cycle occurring during a moment of high technical demand). The participant maintained the setup from Session 3, engaging with the interface without being seated at the piano, as they found this conducive to deep reflection. By the conclusion of the session, the pianist had completed a comprehensive somatic reflection of the entire piece. Finally, we downloaded the performer's annotation notes from the \textit{Breathing Mirror} interface.

\subsection{Analysis}
The first and second authors conducted an inductive thematic analysis, adhering to the reflexive framework proposed by Braun and Clarke~\cite{forbes2022thematic, byrne2022worked}. The first author led the analytical process, beginning with the transcription of audio recordings from Sessions 2 through 4. Given that the interviews were primarily conducted in Japanese, which is the native language of both the interviewer and the pianist, transcriptions were initially generated using the Microsoft Word transcription service. These transcripts were subsequently translated into English via DeepL to facilitate the thematic analysis. Notably, all researchers involved are proficient in English as a working language. To ensure linguistic accuracy of the English version transcriptions, both authors performed a joint audit of the translated documents, cross-referencing them with listening to the original audio to correct technical typos and nuances. The first author then immersed themselves in the data to systematically generate initial codes and candidate themes. The second author facilitated the refinement phase by cross-checking code consistency and reviewing the final themes to ensure they authentically represented the participant's lived experience.

To supplement the qualitative findings, we integrated the annotated timestamps from the session logs, which systematically mapped specific breathing patterns to their corresponding thematic markers within a unified CSV file. These annotations allowed us to triangulate our observations with the primary notes and calculate the frequency of specific breathing-music coupling behaviors. Finally, we generated a timeline visualization to illustrate the distribution and frequency of the six primary breathing patterns observed during the performance across the performance duration.

\section{Findings}
\begin{figure*}[h]
  \centering
  \includegraphics[width=\textwidth]{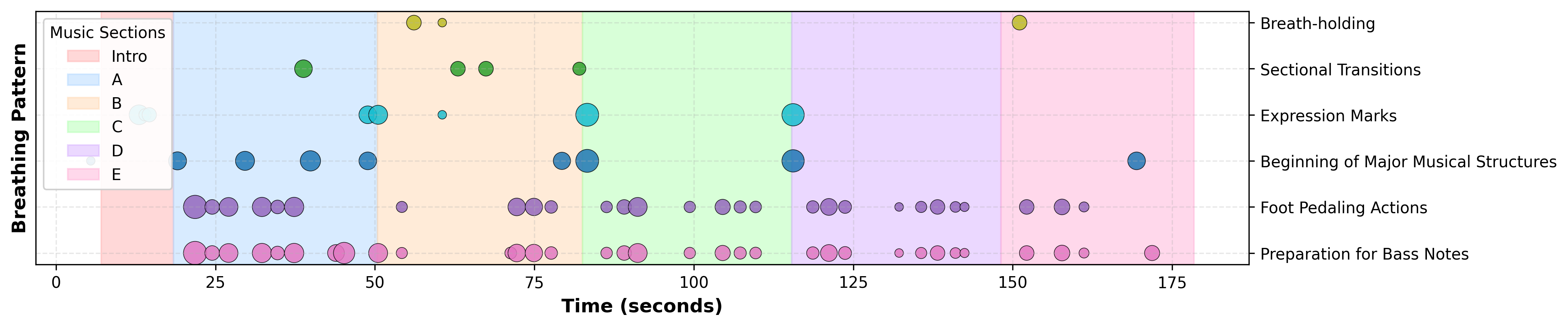}
  \caption{Distribution of breathing patterns observed across musical sections during piano performance. These breathing patterns are categorized by their relationship to specific performance actions: \textit{Breath-holding}, \textit{Sectional Transitions} (preparation for changes in musical scenes), \textit{Expression Marks} (responses to specific score annotations), \textit{Beginning of Major Musical Structures}, \textit{Foot Pedaling Actions}, and \textit{Preparing for Bass Notes}.}
  \Description{Distribution of breathing patterns observed across musical sections during piano performance. These breathing patterns are categorized by their relationship to specific performance actions: Breath-holding, Sectional Transitions (preparation for changes in musical scenes), and Expression Marks (responses to specific score annotations). Additionally, physiological patterns are mapped to physical movements, including New Phrases (the start of major musical structures), Foot Pedaling, and Preparing for Bass Notes.}
  \label{fig:breathing_timeline}
\end{figure*}

\subsection{Timeline Overview}
Our thematic analysis identified six primary themes characterizing the breathing-music relationship, ranked by frequency from top to bottom. To visualize the temporal distribution and density of these behaviors, we generated a timeline event chart (Figure~\ref{fig:breathing_timeline}), mapping each occurrence against the performance duration. By referencing the breathing signal recordings, we scaled the diameter of each circle to be proportional to the amplitude of the breathing cycle.

\subsection{Structural and Harmonic Synchronization}

\begin{figure}[H]
  \centering
    \includegraphics[width=\linewidth]{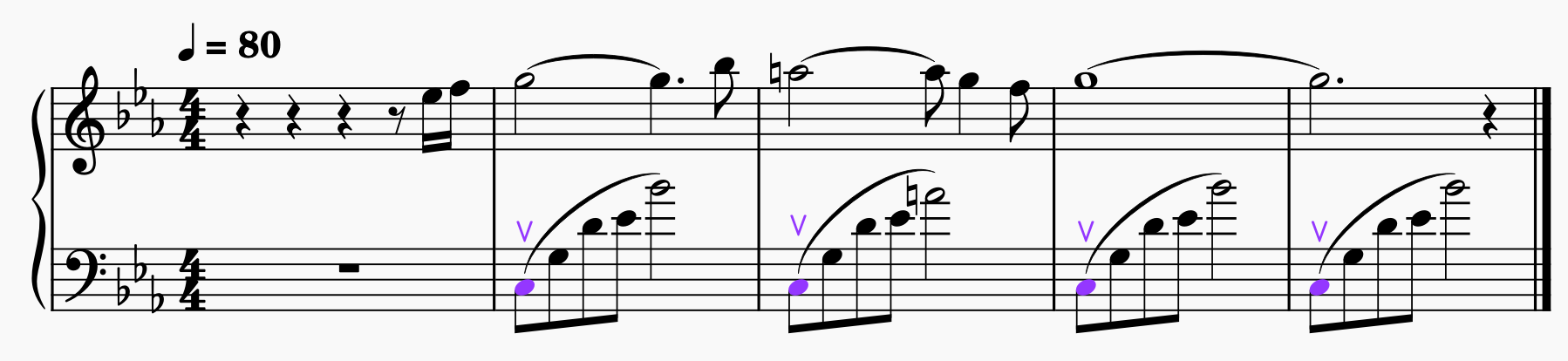}
  \caption{Representative score of the left-hand bass pattern corresponding to the inhalation phase. The purple V markers indicate that the inhalation phase is aligned with the corresponding purple bass notes in the score.}
  \Description{Representative score of the left-hand bass pattern corresponding to the inhalation phase.}
  \label{fig:bassnote_example}
\end{figure}

Many of YK's deep inhalations function as physical preparation for left-hand bass notes, visualized in Figure~\ref{fig:breathing_timeline}, with the intensity of the breath varying according to the surrounding musical annotations. YK acknowledged this as a core habit, stating, ``I really understand my habits that [the inhalation]~\footnote{Square brackets [ ] indicate words added for clarity within YK's quotes; parentheses ( ) are used to provide plain-language explanations for musical terminology.} it's mostly matching the [left hand] bass as the main pattern.'' This relationship is demonstrated in the following Figure~\ref{fig:bassnote_example}, where the bass note highlighted in purple consistently aligns with the rhythmic intake of breath. In sections dominated by arpeggiated left-hand chords (a chord played as a sequence of individual notes), this respiratory preparation aligns with the harmonic foundation. YK explained that ``for the lowest note of an arpeggio, the inhalation prepares for the bass note of the left hand [...] when playing that bass note, I put force in once like this, but for the other notes I'm relaxed.''

The breathing behavior can also emphasize structural transitions and formal phrase boundaries. YK clarified that she adapts her breathing as phrases change, noting that at major section entries, she inhales deeply to establish timing. When a phrase change coincides with a bass note, YK observed that the inhalation seemed to increase: ``Even though the phrase changes, I'm matching it to the bass. [...] But compared to usual, I'm probably inhaling a bit deeper.''

Beyond these structural markers, dynamic breathing patterns emerge in response to changes in mood, atmosphere, or spatial distance. These inhalations act as a bridge between distinct musical sections. For example, ``when entering a new scene,'' YK noted that ``a deep breath is necessary because the mood of the song changes.'' This anticipatory breathing is often precisely synchronized with the onset of new melodic figures. During the interface review, YK identified this coordination: ``Just before ta-ra ta-ra (a recurring rhythmic motif through vocal mimicry) comes in, I'm inhaling once [...] and at the last dot, I start exhaling.''

The most dynamic instances of this expressive breathing occur when the music requires a sense of spatial depth: i.e., YK described visualizing the piano as an orchestral ensemble, where her breath helps project the ``distant'' or ``spatial'' quality of the sound. To achieve this, her inhalations anticipate shifts in instrumental imagery; for example, when the piano's melody transitions to evoke the character of a harp, she uses her breath to prime the body for a distinct change in touch and timbral quality. These shifts often necessitate early or mid-phrase inhalations to shift the atmosphere. 

Finally, YK identified a possible mechanical parallel between respiration and pedaling, suggesting that her inhalation and exhalation cycle mirrors the complex pressing and releasing of the piano pedal to manage phrase clarity: ``the meaning that breathing has is inhaling and exhaling, which is similar to the pedal's pressing and releasing.''

\subsection{Expressive Control and Adaptive Strategies}
\label{subsec:holdbreath}

YK further identified specific musical annotations that trigger breath-holding behaviors. For instance, during expressive markings like ritardando (gradual slowing of the tempo), she adopts a controlled, suppressed breathing style: ``Here is ritardando. [...] Rather than playing vigorously, I might be killing my breath to play carefully.''

\begin{figure}[H]
  \centering
    \includegraphics[width=\linewidth]{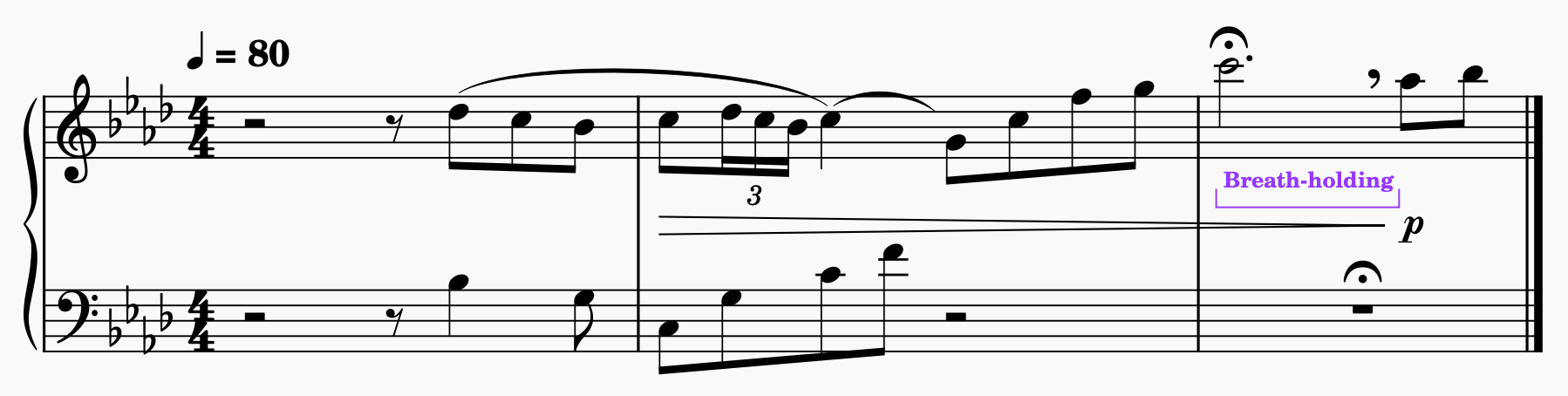}
  \caption{Score example showing the transition into a breath-holding phase.}
  \Description{Score example showing the transition into a breath-holding phase.}
  \label{fig:breathhold_example}
\end{figure}

This intentional suppression is also evident in dynamic passages, where physical exertion necessitates breath control: ``I'm doing it at the limit while holding my breath as much as possible.'' YK describes the process of entering the notes shown in Figure~\ref{fig:breathhold_example}: ``In Japanese, we say `breath-killing' image. There's not much wave [movement], probably [in] the first half, too.'' I want to do that, but from the forte feeling, like after running, it's hard to suppress.'' Beyond purely musical triggers, she attributes these events to her background in performance: ``Maybe it's a habit from wind instrument practices. I used to hold my breath quite a bit.''

In some instances, the length and complexity of a musical phrase impose significant physiological constraints, forcing the pianist to adapt their breathing patterns. To maintain musical flow, she employs respiratory substitution: ``I want to inhale, but the next note is near; instead of inhaling, I exhale.'' This highlights a moment where the performance must shift to accommodate the score's density. As YK explained regarding particularly long melodic lines: ``The main melody is very long, [...] if I match my breathing to the phrase, I'll die [of breathlessness].'' This expression underscores the intense physical limit where the embodied desire to follow a phrase reaches a somatic breaking point. Furthermore, the strict requirements of tempo often preclude standard inhalation entirely: ``It really feels like there's no time to inhale. [...] I might not be breathing. Either breathing very lightly, or not breathing.''

\subsection{Unconsciousness and Consciousness: The Intuition of Breathing}

The relationship between a pianist's subjective awareness and their physiological reality is often characterized by a significant discrepancy. Some unexpected respiratory events are identified where YK's actual breathing patterns differ from her conscious perception. For instance, she noted micro-inhalations used to maintain melodic continuity: ``I'm inhaling just a tiny bit, but this is a phrase that spans two measures. So maybe I don't want to breathe strangely. Like when you're singing yourself, you probably can't breathe [as often as you would like].'' Conversely, some inhalations are more pronounced than anticipated, serving as a physiological ``reset'' for the performer: ``I think I'm breathing more [the amplitude of the breathing cycle] than I thought, and the sensor detects this signal. That [inhalation] is like `now I'm going to play' kind of feeling.''

\begin{figure}[H]
  \centering
    \includegraphics[width=\linewidth]{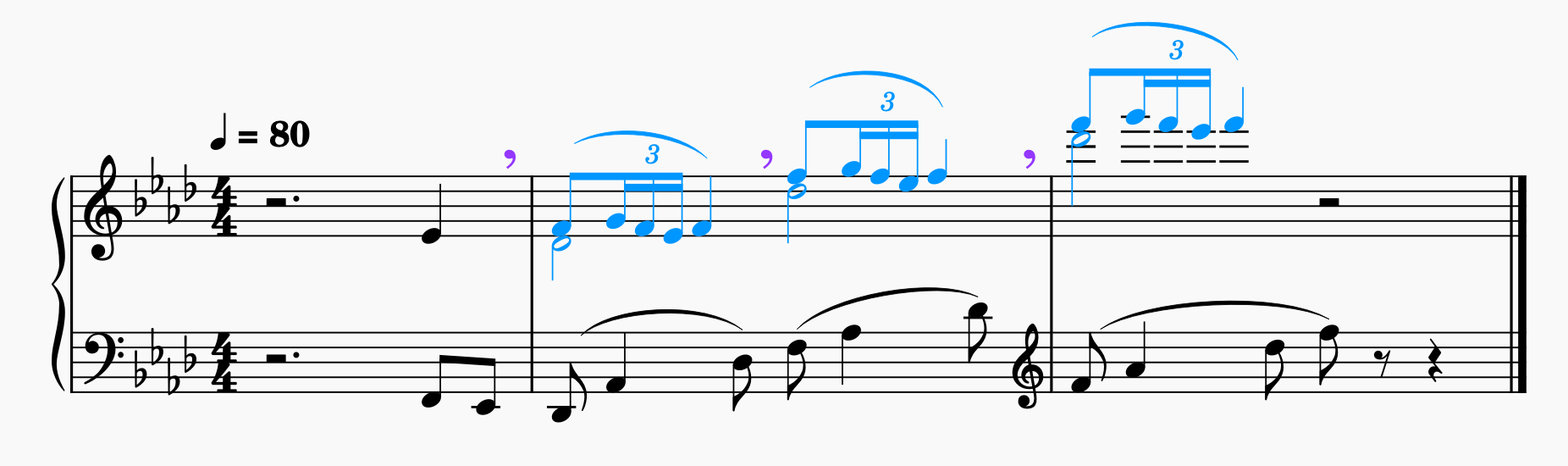}
  \caption{Score example showing inhalation occurred slightly ahead of the corresponding musical event.}
  \Description{Score example showing inhalation occurred slightly ahead of the corresponding musical event.}
  \label{fig:inhaleahead_example}
\end{figure}

This gap in awareness extends to the temporal domain, where YK observed a distinct temporal shift, where her inhalation occurred slightly ahead of the corresponding musical event. This behavior was particularly obvious when approaching high-register passages: ``in the `Ta-ta-ta-ta' (blue notes in Figure~\ref{fig:inhaleahead_example}) part, I was inhaling. [...] It feels like I'm inhaling at the beginning of ``Ta-ta-ta-ta''. When entering the high note, I inhaled around the area just before what I played now, but a bit more ahead.'' While ritardando typically triggers the breath-holding behaviors discussed in Section~\ref{subsec:holdbreath}, the transition back to the primary tempo requires a specific respiratory recalibration. To match the re-entry of the theme, YK noted a ``leisurely'' inhalation synchronized with the structural foundation: ``I'm inhaling once more. After the current place start of the phrase and bass, the main melody phrase begins. [...] To return to the main melody, I think I'm inhaling once more; might be inhaling a bit leisurely to match this bass [notes].''

However, this intuitive respiratory flow is easily disrupted when consciousness is forced toward external stressors or technical errors. Performance-related anxiety significantly influenced YK's respiratory patterns. She identified herself as being particularly susceptible to social and evaluative stress: ``People concern me the most. Having people around. But probably, even if I had simple equipment at home and was told to record, [...] I'd feel stress about recording.'' Furthermore, respiratory deviations were sometimes linked to specific technical challenges or the stress after a performance error. YK noted that the realization of a mistake triggers a reactive change in both her physical execution and her breathing: ``I failed and came out too loud, then I had to keep it loud afterward and adjust a bit. I get quite nervous like that. '' This state of being ``upset'' by an error interrupts the established coordination between breath and structure: ``I'm starting to exhale at the bass. [...] I'm upset about the mistake I made.'' 

These disruptions reveal a blurred boundary between intentional and autonomous respiratory control. When receiving pedagogical instructions regarding breathing's timing and depth, YK occasionally found a discrepancy between the external directive and her intuitive execution. As YK explained: ``Around here I have absolutely no consciousness of breathing. It's not that I'm not breathing, but my consciousness isn't focused on the act of breathing. It's becoming involuntary breathing. But probably, even within involuntary breathing, the timing aligns properly with the musical phrases, so there might still be some consciousness. Consciousness, or rather, unconscious consciousness, the level where consciousness rises from, is probably different.''

Despite this general automaticity, specific musical passages demand fully deliberate respiratory interventions. This is particularly true when a change in sound quality or ``instrumental'' character is required. In these moments, breathing becomes a calculated component of the performance's physical input: ``The phrase entries are still different. [...] It's not related to strong sound, but to the sound quality. It changes suddenly. For example, mimicking the instrument changes [in its orchestra]. At that time, I had to fit how much input based on calculating from what I played before.''

\subsection{Practitioner's Feedback on the Interface Design}

Although this study did not focus primarily on evaluating the wearable device's usability, YK provided positive feedback regarding the breathing belts. When asked about her experience, she emphasized the unobtrusive nature of the hardware: ``I completely forgot I was wearing something.''

Beyond the hardware, YK offered suggestions for the analysis interface after each session. In the initial session, her feedback centered on the general system overview and the pedagogical insights gained from the visualization. She expressed excitement at discovering previously unobserved respiratory patterns, specifically the discrepancy between her perceived and actual behavior, like: ``Looking at this, I think I'm breathing more than I thought.'' 

Based on her feedback, the system evolved from a simple sheet music display to a synchronized MIDI score view to facilitate precise note-to-breath correlation. By the final session, YK provided piano-specific interface refinements. She noted a preference for traditional sheet music over MIDI visualization for better legibility and suggested a modular display to separate the hands: ``I prefer last time, the version with visualizing the sheet music, which feels like it's for people who play piano.''

\section{Discussion}

We propose four emergent somatic themes for evaluation in future research involving larger cohorts of expert pianists.

\subsection{Inhalation as a Somatic Anchor for Structural Bass Notes}
A primary observation in this study is the consistent synchronization between deep inhalation and the onset of left-hand bass notes. As demonstrated in Figure~\ref{fig:breathing_timeline}, the `\textit{Preparation for Bass Notes}' pattern occurs consistently throughout the performance; however, the circle diameters generally exhibit a descending trend in size from Section A through Section E. This indicates that while the somatic pattern is the same, the physiological intensity or breath volume could be various, which suggests that for pianists, breathing can function as a somatic anchor. The act of inhaling provides the trunk stability and postural readiness required for movements used to produce resonant, low-frequency tones. In a broader context, this implies that the respiratory system actively engages in the biomechanics of loading the body to execute structural musical emphasis. This aligns with findings in wind instrument performance, where musicians perform a `long-term reading' of the score to adapt inspiratory volume and muscle recruitment to the upcoming musical demands~\cite{vauthrin2015acoustique}. Future investigations with a larger group of pianists could determine if a correlation exists between the depth of a preparatory breath and the dynamic intensity of the subsequent bass note.

\subsection{Respiratory Suppression as a Proxy for Cognitive and Motor Load}
The phenomenon of ``breath-killing'' indicates a deliberate suppression of physiological `noise', as shown in Figure~\ref{fig:breathing_timeline} by sparse, non-cyclical events rather than repetitive patterns. This sparsity reflects a shift from habitual respiratory cycles to a state of focused motor stillness during high cognitive demand. Notably, the disruption of \textit{Preparation for Bass Notes} and \textit{Foot Pedaling Actions} near these events around at approximately 60--70~s and 150~s, suggests that the performer's ``ideal'' breathing-music coupling is fragile, easily superseded by the cognitive load of error recovery or self-consciousness. By suspending respiration, the performer minimizes extraneous bodily movement to secure the fine motor control necessary for delicate tempo fluctuations. Consequently, breath suppression may serve as an automated metric for identifying sections of peak expressive delicacy. Future cross-participant studies should determine whether this constitutes a universal strategy and identify the specific cognitive or physiological factors to which it relates.

\subsection{Temporal Lead-time in Expressive Anticipation}
The identification of a temporal offset, such as where YK's inhalation occurs slightly ahead of the musical event, highlights the predictive nature of respiration in expert performance. This ``pre-beat'' inhalation is particularly pronounced during shifts in spatial imagery or upper-register melodic entries, suggesting that the breath sets the expressive stage before the first note is even struck. This `short-term anticipation' in respiratory behaviors observed in flutists, who have been shown to adjust lung pressure and muscular positioning in the seconds immediately preceding sound production to match the specific register and nuance of the opening note~\cite{vauthrin2015acoustique}.

From a music technology perspective, understanding this respiratory lead time is crucial for the development of collaborative AI or robotic systems that must anticipate, rather than react to, a human performer's intent. Future cross-participant studies could establish whether the duration and timing of this anticipatory shift are consistent markers of professional training.

\subsection{The Coupling of Respiration and Pedaling}
A unique insight is the functional analogy between the piano's pedagogical use of the pedal and the breathing cycle. The physical act of pressing and releasing the pedal to manage resonance mirrors the inhalation and exhalation cycle used to manage musical phrasing. Beyond individual events, the data reveals that \textit{Preparation for Bass Notes} and \textit{Foot Pedaling Actions} are consistently coupled with time-aligned respiratory cycles throughout the performance. This suggests a deep cross-modal synchronization where the feet and the lungs operate as a unified system to clarify musical textures and prevent the ``blurring'' of harmonic ideas. Investigating this relationship across more pianists could lead to a new understanding of the pianist as a cross-modal synchronization, where breathing, pedaling, and striking may phase-lock to the underlying formal structure of the music.

\section{Conclusion}
This paper explored the intricate relationship between respiration and piano performance through the development and exploration of the \textit{Breathing Mirror}. By utilizing textile-integrated breathing belts and a multiple perspective reflection methodology, we externalized the pre-reflective somatic experiences of a pianist. Our findings demonstrate that the system serves as an inspiring somatic reflection tool. Specifically, we observed synchronization between breathing and harmonic anchors like bass notes and pedaling.

Beyond mere observation, our research identifies breathing as a vital vehicle for structural stability and expressive intent, functioning as a somatic anchor for harmonic transitions and a proxy for cognitive load. Specifically, the synchronization between inhalation and kinematic actions suggests that the pianist's internal biological rhythms may be phase-locked to musical structures. However, further experimental research is required to validate the four emergent somatic themes we proposed concerning the nature of breathing-music coupling.



\section{Ethical Standards}
This study was conducted with the informed consent of the expert pianist, following protocols to ensure participant anonymity and data security. The author utilized a large language model (Gemini) for data visualization and linguistic refinement after drafting the code and manuscript. All AI-assisted outputs were manually verified and edited by the authors.

\begin{acks}
This work was conducted during the Yamaha Global Internship Program and was supported by an FRQSC Doctoral Scholarship (No. 0000342567) from the Fonds de recherche du Québec. The authors would like to express their sincere gratitude to Katsunori Suzuki, Naoto Kojima, and other related colleagues for their essential technical support in the development and maintainance of the breathing belts.

We also thank the colleagues at Yamaha who provided valuable feedback during the first author's internal presentations, with special thanks to Jingwei Zhao, Jian Gao, Angela Zhang, Xinmeng Luan, Chuen Ken Ng, and Rei Nishiyama. We also appreciate the insightful discussions with Kanyu Chen, Theodora Nestorova, and Hanwen Zhang regarding breathing in singing and piano performances. 
\end{acks}

\bibliographystyle{ACM-Reference-Format}
\bibliography{sample-references}


\end{document}